\documentclass[prl,twocolumn,showpacs,superscriptaddress,amsfonts,amsmath,floatfix]{revtex4-1}

\usepackage{graphicx}
\usepackage{color}

\begin{document}

%=q======================================================================================
\title{Synthetic Lorentz force in classical atomic gases
via Doppler effect and radiation pressure}

\author{T.~Dub\v{c}ek}
\affiliation{Department of Physics, University of Zagreb, Bijeni\v{c}ka c. 32, 10000 Zagreb, Croatia}
\author{N. \v{S}anti\'{c}}
\affiliation{Department of Physics, University of Zagreb, Bijeni\v{c}ka c. 32, 10000 Zagreb, Croatia}
\author{D.~Juki\'{c}}
\affiliation{Department of Physics, University of Zagreb, Bijeni\v{c}ka c. 32, 10000 Zagreb, Croatia}
\affiliation{Max Planck Institute for the Physics of Complex Systems, N\"{o}thnitzer Str.
38, 01187 Dresden, Germany}
\author{D.~Aumiler}
\affiliation{Institute of Physics, Bijeni\v{c}ka c. 46, HR-10000 Zagreb, Croatia}
\author{T.~Ban}
\affiliation{Institute of Physics, Bijeni\v{c}ka c. 46, HR-10000 Zagreb, Croatia}
\author{H.~Buljan}
\email{hbuljan@phy.hr}
\affiliation{Department of Physics, University of Zagreb, Bijeni\v{c}ka c. 32, 10000 Zagreb, Croatia}

\date{\today}

\begin{abstract}
We theoretically predict a novel type of synthetic Lorentz force 
for classical (cold) atomic gases, which is based on the Doppler effect 
and radiation pressure. 
A fairly spatially uniform and strong force can be constructed for gases in 
macroscopic volumes of several cubic millimeters and more. 
This opens the possibility to mimic classical charged gases in  
magnetic fields in cold atom experiments. 
\end{abstract}

\maketitle
%\narrowtext
%\newpage

The quest for synthetic magnetism in quantum degenerate atomic gases is motivated 
by producing controllable quantum emulators, which could mimic complex quantum 
systems such as interacting electrons in magnetic fields \cite{Bloch2012}. An appealing idea is 
to place the atomic gas in a specially tailored laser field which, due to laser-atom 
interactions, acts as a synthetic magnetic field for neutral atoms \cite{Dal2011}. 
The mechanism is based on the analogy between the Aharonov-Bohm phase accumulated 
when a charged quantum particle undergoes a closed loop in a magnetic field, and the Berry 
phase accumulated when an atom adiabatically traverses a closed loop in the tailored 
laser field \cite{Dum1996,Dal2011}.

Recent experiments in bulk Bose-Einstein condensates (BECs) have produced 
synthetic magnetic fields by spatially dependent optical coupling between the 
internal states of the atoms \cite{Lin2009,LeB2012}. 
Superfluid vortices \cite{Lin2009} and the Hall effect \cite{LeB2012} 
were observed as signatures of synthetic magnetism in those BECs. 
Synthetic magnetism in optical lattices is achieved by engineering 
the complex tunneling parameter between the lattice sites, which 
is experimentally accomplished by different means \cite{Aidelsburger2011,Struck2012}. 
Interestingly, even Dirac monopoles were observed in a synthetic 
magnetic field produced by a spinor BEC \cite{Ray2014}. 
Synthetic magnetic fields for light (e.g., see \cite{Carusotto2013}) are also 
attractive. Recently they were observed in deformed honeycomb 
photonic lattices \cite{Rechtsman2013}. Noninertial effects 
were studied in rotating waveguide arrays \cite{Jia2009}.

However, {\em classical} (rather than quantum degenerate) cold atomic gases have 
been circumvented in the quest for synthetic magnetism, even though they could 
emulate in a controllable fashion, and in table-top experiments, 
versatile complex classical systems (e.g., see \cite{Struck2011,Baudouin2014}); 
one desirable system for table-top emulation is tokamak plasma. 
We emphasize that here we consider classical atomic gases. 
This differs from using quantum degenerate gases to mimic frustrated classical 
magnetism in Ref. \cite{Struck2011}.  
Laser forces on atoms in classical gases can generally depend on atomic velocity 
\cite{Chu1985} and position \cite{Met1999}. 
A typical example is the Doppler cooling force - a viscous 
damping force that cools a classical gas to $\mu$K temperatures \cite{Chu1985,Met1999}. 
Here we demonstrate a novel scheme for creating synthetic Lorentz force via the Doppler 
effect and radiation pressure, which is applicable for classical cold atomic 
gases. The experimental realization of the scheme is proposed with $^{87}$Rb atoms 
cooled in a Magneto-Optical Trap (MOT). 
The signature of the Lorentz force can be observed in the motion 
of the center of mass (CM) and/or the shape of the atomic cloud.

Numerous schemes have been proposed to create synthetic magnetic fields 
with ultracold atoms (see \cite{Dal2011,Bloch2012,Cooper2008} for reviews). 
In the approach based on the Berry phase \cite{Dum1996}, when atoms 
move in space, they adiabatically follow the ground state of the light-atom 
coupling (dressed state), which depends on the spatial coordinates \cite{Dal2011,Dum1996}. 
Their CM wavefunction acquires a geometric (Berry) 
phase, which corresponds to the gauge potentials \cite{Dal2011}. The synthetic magnetic 
(and electric \cite{Lin2011}) fields are derived from these gauge potentials \cite{Dal2011}. 
In these schemes spontaneous emission must be minimized to prevent heating of 
the ultracold gas. For this reason, the dressed (ground) state is often a superposition of 
quasidegenerate ground states \cite{Dum1996,Juzeliunas2004,Juzeliunas2006}, i.e., 
the population of excited states is negligible \cite{Juzeliunas2004,Juzeliunas2006}. 
A semi-classical interpretation of geometric gauge potentials, i.e., 
the connection with the Lorentz force was reported in Ref. \cite{Cheneau2008}.

Another avenue for creating artificial magnetic fields in ultracold atomic gases 
is to rotate the system at some angular frequency, 
which provides the synthetic Lorentz force in the rotating frame \cite{Cooper2008}; 
the role of the Lorentz force is played by the Coriolis force. 
This scheme is suitable for rotationally invariant trapping potentials. 
However, the laser-atom interactions avenue is more 
appealing since it does not impose symmetries 
and produces synthetic magnetic fields in the laboratory frame \cite{Dal2011}.

In classical atomic gases, 
any scheme for synthetic magnetism must be operational 
on atoms moving with fairly large velocities (at least up to $\sim 0.5$~m/s). 
The Berry phase method demanding adiabatic dynamics is therefore limited \cite{Dal2011}. 
On the other hand, schemes for classical gases do not need to be limited by 
avoiding spontaneous emission. 
Next, classical gases in a standard MOT are typically of millimeter size \cite{Met1999} and 
the synthetic Lorentz force should therefore be large in volumes of at 
least a few cubic millimeters. 
With these guidelines in mind, it seems prosperous to seek for a novel scheme using laser-atom 
interactions for creating synthetic Lorentz forces in classical atomic gases.

\begin{figure}%[htbp]
\centerline{
\mbox{\includegraphics[width=0.47\textwidth]{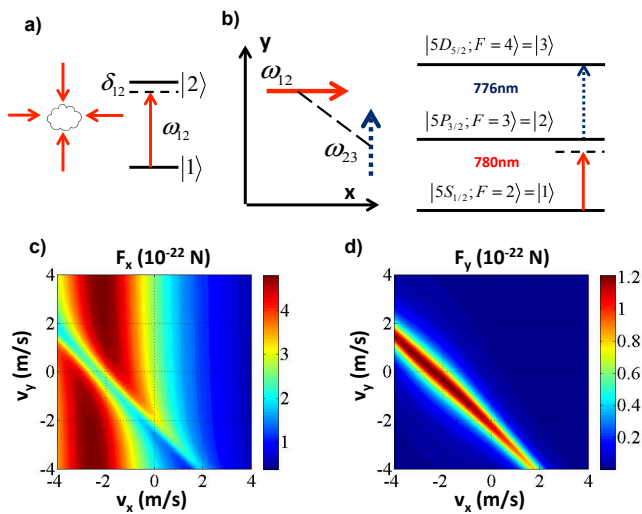}}
}
\caption{(Color online)
Sketch of the main idea for constructing the 
synthetic Lorentz force. (a) Illustration of the setup for the 
standard Doppler cooling force (using two-level atoms).
(b) The idea for the synthetic Lorentz force in the simplest 
three-level system that can be realized with $^{87}$Rb atoms. 
Dashed line indicates that two-step absorption of 
$\omega_{12}+\omega_{23}$ yields $F_y$. 
The force components $F_x$ (c) and $F_y$ (d) calculated 
as a function of the atomic velocity. 
See text for details.}
\label{idea}
\end{figure}

The scheme proposed here is based on the Doppler effect and radiation 
pressure. The standard Doppler cooling force arises 
when a laser field is red-detuned compared to the atomic resonance frequency as 
sketched in Fig. \ref{idea}(a) \cite{Chu1985,Met1999}. 
Due to the Doppler effect, the atom has greater probability for absorbing a 
photon when it moves towards the light source. 
Absorption changes the atom's momentum along the laser propagation axis, whereas 
spontaneously emitted photons yield random kicks. Cycles of 
absorption and emission result in a viscous damping force 
${\bf F}_{D}({\bf v}) \approx -\alpha{\bf v}$ for small velocities \cite{Met1999}; 
this force is collinear with the velocity and is used to obtain optical molasses 
\cite{Met1999}. 

Our first objective is to construct a laser-atom system (in the $xy$ plane) 
where $F_y$ depends on $v_x$. 
To achieve this via Doppler effect we utilize the multilevel structure of atoms. 
The simplest scheme is sketched in Fig. \ref{idea}(b), where a three-level atom 
interacts with two orthogonal laser beams (linearly polarized along $z$). 
The laser $\omega_{12}$ is red detuned: $\delta_{12}=\omega_{12}-(E_2-E_1)\hbar^{-1}<0$, 
whereas $\omega_{23}$ is on resonance: $\delta_{23}=0$. 
The absorption of $\omega_{23}$ photons, which results in $F_y$, is the second 
step in the two-step two-photon absorption process: 
$|1 \rangle \rightarrow |2 \rangle \rightarrow|3 \rangle$. 
The probability for the two-step absorption depends on 
the Doppler shifted detuning values $\delta_{12}-k_{12}v_x$ and 
$\delta_{23}-k_{23}v_y$, which provides the desired dependence of $F_y$ on $v_x$. 
The maximum in $F_y$ is expected for atoms with velocity 
$(v_x=\delta_{12}/k_{12},v_y=\delta_{23}/k_{23})$, i.e., when each 
of the two steps is resonant.

The force can be calculated by using density matrices and the Ehrenfest theorem 
as described in detail in Ref. \cite{Met1999}. 
First we (numerically) solve the optical Bloch equations to find 
the stationary density matrix $\hat \rho$ 
for an atom with velocity ${\bf v}$; the matrix elements are 
$\rho_{ij}=\sigma_{ij} e^{i \omega_{ij} t}=\rho_{ji}^*$, and $d \sigma_{ij} /dt=0$;
$\omega_{ij}$ is the frequency of the laser driving the transition $|i\rangle \rightarrow |j\rangle$. 
In the calculation, the following parameters are used \cite{Met1999}: 
the energies $E_j$ of the levels participating in the interaction ($j=1,\ldots,N$), 
the Rabi frequencies $\Omega_{ij}$, 
detuning values $\delta_{ij}$, 
the wavevectors ${\bf k}_{ij}$ of the lasers, 
and the decay parameters of the excited states  
($\Gamma_{ji}$ is the decay rate via $|j\rangle \rightarrow |i\rangle$; 
the total width of state $|j\rangle$ is $\Gamma_j=\sum_{i<j} \Gamma_{ji}$). 
The force is given by 
${\bf F}=\langle - \nabla_r \hat H \rangle = - {\mbox Tr}(\hat \rho \nabla_r \hat H)$, 
where $\hat H$ is the Hamiltonian associated with the dipole interaction,
and $\nabla_r={\bf \hat x}\partial/\partial x+{\bf \hat y}\partial/\partial y$ \cite{Met1999}.
For plane (traveling) waves used here, 
${\bf F}=-\sum_{i=1}^{N-1} \sum_{j=i+1}^{N} \hbar {\bf k}_{ij} \text{Im} ( \sigma_{ij} \Omega_{ij}^*)$. 
The density matrix depends on the Doppler shifted detuning values 
$\delta_{ij}-{\bf k}_{ij} \cdot {\bf v}$, which provides the velocity dependence of the 
force \cite{Met1999}.

It should be emphasized that the ideas for constructing synthetic 
Lorentz forces presented here are general and potentially applicable to various atomic species. 
For concreteness, the ideas are presented for $^{87}$Rb atoms using experimentally 
relevant atomic states and transitions. 
The three-level system that can be used to 
experimentally realize the simplest scheme is presented in Fig. \ref{idea}(b). 
The transition wavelengths are $\lambda_{12}=780$~nm \cite{Steck} and $\lambda_{23}=776$~nm 
\cite{Sheng2008}. 
The decay rate of the $|5P_{3/2}\rangle $ hyperfine states is $\Gamma_P=2\pi\times 6.1$~MHz \cite{Steck}, 
and $\Gamma_D=2\pi\times 0.66$~MHz for $|5D_{5/2}\rangle$ states \cite{Sheng2008};
the decay pattern is $\Gamma_{32}=\Gamma_D$, $\Gamma_{31}=0$, and $\Gamma_{21}=\Gamma_P$ \cite{Steck,Sheng2008}. 
In Fig. \ref{idea}(c,d) we illustrate ${\bf F}({\bf v})$ for detuning values 
$\delta_{12}=-0.5\Gamma_P$, $\delta_{23}=0$, and Rabi frequencies 
$\Omega_{12}=0.12\Gamma_P$, and $\Omega_{13}=0.34\Gamma_P$. 
As expected, the maximum of the force $F_y$ occurs when $v_x=\delta_{12}/k_{12}$ and 
$v_y=\delta_{23}/k_{23}$. Interestingly, $F_y(v_x,v_y)$ has the shape of a 
mountain ridge peaked at $\delta_{12} + \delta_{23} - k_{12} v_x- k_{23} v_y=0$. 
This is a consequence of the fact that the intermediate state 
$|2\rangle$ is much broader than state $|3\rangle$. 
For the two-step absorption to be effective, the Doppler shifted detuning of the 
first photon should roughly be $|\delta_{12} - k_{12} v_x| < \Gamma_2$,  
and the total detuning $|\delta_{12} + \delta_{23} - k_{12} v_x - k_{23} v_y| < \Gamma_3$; 
since $\Gamma_{3}\ll \Gamma_{2}$, the velocities satisfying these inequalities 
are close to the ridge line.
The ridge can be shifted in the $v_x v_y$ plane by 
changing the detuning values. The scheme above illustrates the 
main idea towards constructing the synthetic Lorentz force via the Doppler effect.

Note that the force in the $x$-direction is also altered for atoms with velocities 
at the ridge. The presence of second step transition $|2\rangle\rightarrow |3\rangle$ 
changes the populations of all three levels, which affects the rate of 
first step transition $|1\rangle\rightarrow |2\rangle$ and hence $F_x$. 
It should be noted that deformations of the ridge can arise for larger 
Rabi frequencies due to the Autler-Townes effect \cite{AutlerTownes}.

In order to provide a general framework for our sheme we 
Taylor expand the force in velocity up to the linear term:
\begin{eqnarray}
\begin{bmatrix}
F_x \\
F_y
\end{bmatrix}
& = &
\begin{bmatrix}
F_{x0} \\
F_{y0}
\end{bmatrix}
+
\begin{bmatrix}
\alpha_{xx} & 0 \\
0 & \alpha_{yy}
\end{bmatrix}
\begin{bmatrix}
v_x \\
v_y
\end{bmatrix}
+
\begin{bmatrix}
0 & \alpha_{xy} \\
\alpha_{yx} & 0
\end{bmatrix}
\begin{bmatrix}
v_x \\
v_y
\end{bmatrix}
\nonumber
\\
\nonumber 
\\
& = & 
{\bf F}_0+{\bf F}_{D}({\bf v})+{\bf F}_{SL}({\bf v}).
\label{liner}
\end{eqnarray}
Here $\alpha_{ij}=\partial F_i/\partial v_j$ evaluated at ${\bf v}=0$ ($i,j\in \{ x,y \}$). 
This form is often an excellent approximation for ${\bf F}({\bf v})$  because 
of the small velocities of cold atoms. 
The third term ${\bf F}_{SL}({\bf v})$ is a general form of the synthetic 
Lorentz force with components perpendicular to the velocity components:
$F_{SL,x}=\alpha_{xy}v_y$, $F_{SL,y}=\alpha_{yx}v_x$ \cite{SLdefinition}. 
The force on a standing atom is ${\bf F}_{0}$; 
the components of the standard Doppler force are 
$F_{D,x}=\alpha_{xx}v_x$, and $F_{D,y}=\alpha_{yy}v_y$.  
When $\alpha_{xy}=-\alpha_{yx}$, ${\bf F}_{SL}$ takes the 
form of the standard Lorentz force: 
${\bf F}_{SL}={\bf v}\times {\bf B}^*$, where ${\bf B}^*=\alpha_{xy} {\bf \hat z}$ 
\cite{SLdefinition}.
\begin{figure}%[htbp]
\centerline{
\mbox{\includegraphics[width=0.470\textwidth]{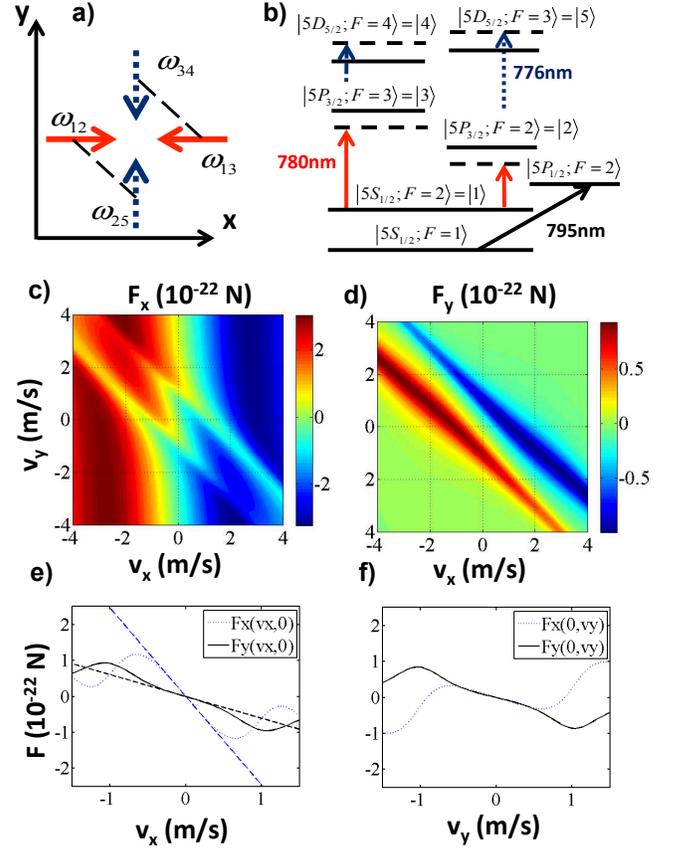}}
}
\caption{(Color online) 
The five-level scheme. (a) 
The symmetric configuration of lasers for creating the synthetic Lorentz 
force, and (b) its realization with hyperfine $^{87}$Rb levels (b). 
Density plots of $F_x$ (c), and $F_y$ (d) as a function of velocity. 
Cross sections $F_x(v_x,0)$ and $F_y(v_x,0)$ (e), and 
$F_x(0,v_y)$ and $F_y(0,v_y)$ (f). 
See text for details.}
\label{5level}
\end{figure}

Let us illustrate a few force patterns ${\bf F}({\bf v})$ that can be achieved 
with our scheme. Consider a system of five-level atoms and two orthogonal pairs of 
counter-propagating beams depicted in Fig. \ref{5level}(a). 
This is simply a generalization of the idea presented in Fig. \ref{idea} 
with a symmetric pair of two-step arms such that ${\bf F}_0=0$. 
It can be experimentally realized by using hyperfine levels of 
$^{87}$Rb depicted in Fig. \ref{5level}(b);  
the use of re-pumper laser is mandatory since the chosen 
five-level system is not closed: 
$|5P_{3/2};F=2\rangle \xrightarrow{50\%} |5S_{1/2};F=1\rangle=|0\rangle
\xrightarrow{795nm}|5P_{1/2};F=2\rangle=|6\rangle \xrightarrow{50\%}
|5S_{1/2};F=2\rangle$ (this is included in our calculations). 
The Rabi frequencies and detuning values for the transitions are 
$\Omega_{12}=\Omega_{13}=0.11\Gamma_P$, and $\delta_{12}=\delta_{13}=-0.5\Gamma_P$. 
The pairs of beams along $x$ are red detuned, while the pairs along $y$ are 
blue detuned (by a smaller magnitude): $\delta_{25}=\delta_{34}=0.25\Gamma_P$; 
$\Omega_{25}=\Omega_{34}=0.38\Gamma_P$. 
The repumper is on resonance with high intensity $\Omega_{06}=1.77\Gamma_P$, 
in a standing wave configuration (it does not produce net force on atoms). 
The decay pattern is given by 
$\Gamma_{21}=\Gamma_{20}=0.5\Gamma_P$, 
$\Gamma_{31}=\Gamma_P$, 
$\Gamma_{43}=\Gamma_D$, 
$\Gamma_{53}=0.2\Gamma_D$, $\Gamma_{52}=0.8\Gamma_D$, 
and $\Gamma_{60}=\Gamma_{61}=0.5\Gamma_P$;
the rest of $\Gamma_{ji}=0$. 
In Fig. \ref{5level}(c,d) we show the force ${\bf F}({\bf v})$. 
Atoms moving towards the left (right) will experience $F_y>0$ ($F_y<0$, respectively). 
From Fig. \ref{5level}(e,f) we see that the force depends linearly 
on the velocity $v_x$ for velocities below $\sim 0.7$ m/s (which includes essentially 
all atoms in a standard $^{87}$Rb MOT \cite{Met1999}). 
The two ridges in $F_y$ correspond to the pair of arms of the two-step absorption; 
their position and shape was explained in Fig. \ref{idea}(d). 
By changing the detuning values, the ridges can be shifted in the $v_xv_y$ plane, 
which changes the parameters $\alpha_{ij}$ and therefore the strength of the synthetic 
Lorentz force. 
\begin{figure}%[htbp]
\centerline{
\mbox{\includegraphics[width=0.48\textwidth]{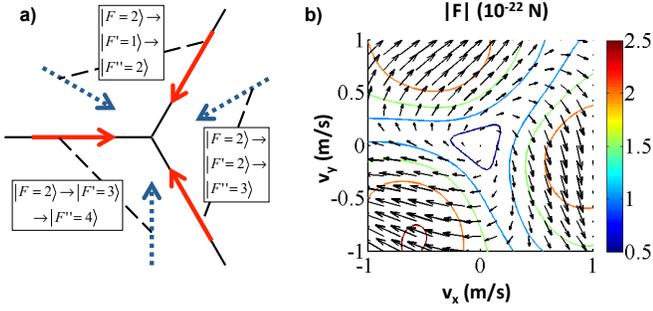}}
}
\caption{(Color online)
The tripod configuration of the three two-step excitations arms, 
and the obtained force. (a) Red solid arrows depict first-step excitations 
(red detuned), and blue dotted arrows depict second steps (blue detuned).
Black dashed lines connect beams that correspond to one arm;
$|F=2\rangle$ denotes the $5S_{1/2}$ hyperfine state, 
$|F'=1,2,3\rangle$ denote three $5P_{3/2}$ hyperfine states, 
and 
$|F''=2,3,4\rangle$ denote three $5D_{5/2}$ hyperfine states.
(b) Contour lines and length of the arrows correspond to the magnitude of 
the force $|{\bf F}({\bf v})|$.} 
\label{vxB}
\end{figure}

It should be noted that because our approach is 
based on the Doppler effect, it usually also yields the 
Doppler (cooling) force ${\bf F}_D$.
If for some reason this is not wanted, dissipation can be diminished 
(for example by using one blue and one red detuned laser in the 
counterpropagating configuration for the first step excitation). 
Moreover, the synthetic force can be made of the form ${\bf v}\times {\bf B}^*$: 
By using three arms of the two-step scheme at $120^\circ$ [Fig. \ref{vxB}(a)], 
one can obtain the force plotted in Fig. \ref{vxB}(b). 
Two arms are identical as in Fig. \ref{5level}(b), and the third arm is 
$|5S_{1/2};F=2\rangle \rightarrow |5P_{3/2};F=1\rangle \rightarrow |5D_{5/2};F=2\rangle $. 
The Rabi frequency of the first (second) step in all arms is 
$0.11\Gamma_P$ ($0.77\Gamma_P$); the detuning values are 
$-0.5\Gamma_P$ ($0.25\Gamma_P$) for the first (second) step. 
Clearly, the force rotates around zero in the $v_xv_y$ plane. 
Strictly, the force field is invariant under rotation by 120 degrees, 
however, for small velocities it is effectively rotationally invariant. 
By fitting ${\bf F}(\bf v)$ to Eq. (\ref{liner}) we 
obtain $\alpha_{xy}=-\alpha_{yx}=0.23\times 10^{-21}$~Ns/m, i.e., ${\bf B}^*=\alpha_{xy} {\bf \hat z}$.
The cyclotron frequency for $^{87}$Rb atoms corresponding to our forces is  
$\alpha_{xy}/m \approx 1.6$~kHz.
It should be emphasized that, because we are using hyperfine levels of $^{87}$Rb, 
the scheme can be achieved with two CW lasers at 
780~nm and 776~nm by using acoustic optical modulators (AOMs), 
i.e., it is experimentally viable.

The prediction of the synthetic Lorentz force is made for individual atoms, 
however, we should propose its signature in the CM motion and/or shape of a cold 
atomic cloud containing a huge number (say $\sim 10^9$ \cite{Met1999}) of atoms. 
To this end we propose a quench-type scenario(s). 
First, we assume that an atomic cloud is present in the MOT, 
and cooled to mK-$\mu$K temperatures.  The laser fields driving the MOT have 
much larger Rabi frequencies than lasers producing synthetic Lorentz force. 
The latter will slightly heat up the cloud, but will not change its shape.  
Then, at $t=0$, the MOT lasers and the magnetic field 
are suddenly turned off (it can be done within less than 1~$\mu$s, which is 
essentially instantaneous for this system). After $t=0$, the cloud starts moving 
in the presence of the synthetic Lorentz and Doppler forces. 
We focus on dynamics in the $xy$ plane. Moreover we assume that gravity 
is in the $z$ direction and does not influence observations. 
The typical experimental observation time for the measurements proposed 
here is 5-10~ms; an initially standing atom will fall for 
0.12-0.49 mm. The laser fields creating the synthetic Lorentz forces can be made 
of much larger diameter (on the order of several cm). 
Since dynamics in the $xy$ plane is independent of the dynamics in the 
$z$-plane, we do not expect a significant influence of gravity on our predictions below. 

\begin{figure}%[htbp]
\centerline{
\mbox{\includegraphics[width=0.48\textwidth]{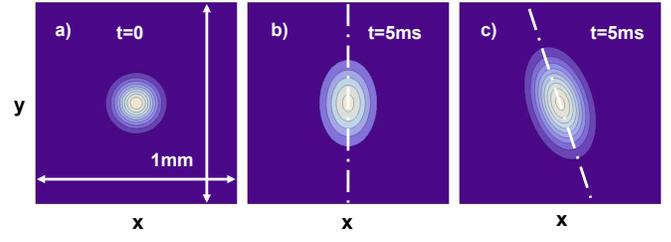}}
}
\caption{(Color online)
Dynamics of the shape of the cloud during expansion 
in the presence of synthetic Lorentz force ${\bf F}_{SL}({\bf v})$ 
and/or Doppler force ${\bf F}_{D}({\bf v})$. 
The force corresponds to that from Fig. \ref{5level}. 
(a) Density of the cloud at $t=0$, 
(b) after $5$~ms of expansion under the action of ${\bf F}_{D}({\bf v})$, 
(c) and after $5$~ms expansion in the presence of ${\bf F}_{D}({\bf v})+{\bf F}_{SL}({\bf v})$. 
See text for details.}
\label{figFP}
\end{figure}

We will discuss two scenarios for the force plotted in Fig. \ref{5level}. 
First, if the cloud is given an initial velocity ($v_x=0.15$~m/s,$v_y=0$), 
it will move along $x$ due to inertia, but its CM will also move in the negative 
$y$ direction due to the synthetic Lorentz force. 
After 10~ms the shift in $y$ is $\sim 0.2$~mm, which is observable in MOT experiments. 
Initial velocity can be achieved by inducing oscillations of the cloud in the MOT trap 
for $t<0$ (e.g., see \cite{Xu2002}). 

Second we discuss expansion of the cloud by employing the Fokker-Planck equation \cite{Met1999}:
\begin{equation}
\frac{\partial P({\bf x},{\bf v},t)}{\partial t}+ {\bf v} \cdot \nabla_r P=
\frac{-1}{m}\nabla_v \cdot [({\bf F}_D+{\bf F}_{SL})P]+
\frac{D}{m^2} \nabla_v^2 P.
\label{Fokker}
\end{equation}
Here, $P({\bf x},{\bf v},t)$ is the distribution of particles in the 
phase space; $D$ is the diffusion constant, approximately given 
by $D \approx (\hbar k)^2 \sum_j \rho_{jj} \Gamma_j$ \cite{Met1999}, where $k\approx 2\pi/780$~{nm}$^{-1}$;
$\nabla_v={\bf \hat x}\partial/\partial v_x+{\bf \hat y}\partial/\partial v_y$.
For forces ${\bf F}_D+{\bf F}_{SL}$ linearized in velocity (\ref{liner}), 
the Fokker-Planck equation is solved by the ansatz:
\begin{equation}
P(x,y,v_x,v_y,t)=P_0 \exp \{ - \sum_{ij=1}^{4} \frac{1}{2} a_{ij}(t) \eta_i \eta_j \},
\label{ansatz}
\end{equation}
where $(\eta_1,\eta_2,\eta_3,\eta_4)=(x,y,v_x,v_y)$; after inserting (\ref{ansatz})
in Eq. (\ref{Fokker}), one obtains 10 coupled 
ordinary differential equations (ODEs) for the functions $a_{ij}(t)$; 10 because 
$a_{ij}=a_{ji}$ by construction. 
These coupled ODEs are solved numerically and the results are 
plotted in Fig. \ref{figFP} for the following parameters: 
$(\alpha_{xx},\alpha_{xy},\alpha_{yx},\alpha_{yy})=-(2.4,0.60,0.69,0.58)\times 10^{-22}$~Ns/m, 
and $D/m_{Rb}^2=31$~m$^2$s$^{-3}$; the initial state is 
$P=P_0 \exp \{-({\bf x}/x_0)^2-({\bf v}/v_0)^2 \}$, where $x_0=1$~mm, and $v_0=0.25$~m/s.  
Starting from a centrosymmetric cloud plotted in Fig. \ref{figFP}(a), 
in the presence of solely the Doppler force, the cloud expands asymmetrically 
[Fig. \ref{figFP}(b)] because $|\alpha_{xx}|>|\alpha_{yy}|$. 
The signature of the synthetic Lorentz force is the rotation of 
the asymmetric cloud in the $xy$ plane during expansion [see Fig. \ref{figFP}(c)]. 
The interpretation is simple: particles moving to the left (right) are pushed up (down),  
as can be inferred from Fig. \ref{5level}(d). 
There is another effect: the change in $F_y$ for a given atomic velocity group
also changes $F_x$ for that group, as discussed above. 
For the parameters corresponding to Figs. \ref{5level} and \ref{figFP}, 
besides the targeted $\alpha_{yx}<0$, we incidentally also obtained 
$\alpha_{xy}<0$ (for small velocities). 
Note that expansion in the rotationally symmetric force field presented in 
Fig. \ref{vxB} would cause rotation of atoms around the center, but this would not be visible 
in the density (in the proposed scenario it is essential to have 
$|\alpha_{xx}| \neq |\alpha_{yy}|$). By shining a red detuned laser 
beam in the plane of such a rotationally invariant but rotating cloud, one would have 
different absorption in the part of the cloud moving towards (away) 
from the laser beam due to the Doppler effect; this seems like one viable scheme 
to observe rotation of the cloud.

Before closing, let us discuss specific approximations that we used
here to simplify the discussion. 
First, we neglected the absorption of the lasers in the cold atomic cloud. 
Absorption changes intensity of beams across the cloud, and therefore 
introduces spatial dependence of the synthetic 
Lorentz force (and not only the velocity dependence). This effect can be 
reduced by using clouds with lower density (say $10^9$ atoms per cm$^3$), 
or by using lasers with higher intensity (closer to saturation). 
The latter approach will also increase the diffusion coefficient.
Second, in our proposal we neglected the Zeeman structure of the atomic levels. 
This simplification is acceptable when dynamics of the cloud 
does not occur in a magnetic field (i.e., Zeeman splitting is absent), 
as in the two scenarios described above. 
Next, the dipole moments of different transitions used in the scheme 
will be generally different (they also depend on the polarization of the light used). 
The key goal one has to achieve is to have the same Rabi frequencies 
for all first (second) steps in each arm as in our examples above. 
In experiments, this can be realized by using light 
of different intensity in steps that have different transition dipole
moments. This could in principle be achieved by balancing the forces 
arising from different arms. 
Finally, let us note that the internal dynamics occurs on a much faster 
time-scale than CM motion; the bottleneck for internal dynamics is the 
lifetime of the $5D_{5/2}$ state of 240~ns, whereas the typical time-scale for CM motion 
is 1~ms.

In conclusion, we have demonstrated a scheme for creating synthetic Lorentz forces 
in cold classical atomic clouds, based on the Doppler effect and radiation pressure.
We envision that following these ideas, one could design cold gas experiments 
to mimic classical charged gases in magnetic fields. 
One desired classical system for emulation is tokamak plasma. 
A necessary (but not sufficient) step towards this goal is to have 
a scheme for producing synthetic magnetic fields for classical gases. 
The next step towards mimicking tokamak plasma would be to 
construct a toroidal synthetic magnetic field, which is beyond the scope of this paper. 
Here we have predicted synthetic Lorentz forces of magnitude 
$F/v \approx 0.23\times 10^{-21}$~Ns/m in macroscopic volumes of a few mm$^3$ and more.
The maximal volume depends on the intensities of lasers; with standard diode 
lasers one could achieve the synthetic Lorentz force in at least 1~cm$^3$. 
As a reference point we note that the obtained force on a unit charge 
particle (of any mass) is produced by a magnetic field of 1.5~mT. 
The cyclotron frequency (which includes the particle mass) for $^{87}$Rb atoms 
corresponding to our forces is $F/(mv)\approx 1.6$~kHz, 
which is large enough to see the phenomena associated to the 
synthetic Lorentz force on the time-scale of envisioned experiments. 
Even stronger forces can be achieved for larger intensities of the lasers 
at the expense of more heating and diffusion. 
We envision that our concept involving two-photon 
absorption could be applicable in other systems, e.g., for suspended 
nanoparticles with nonlinear index of refraction where 
one laser beam would induce index change, and thus influence the force of another 
(say perpendicular) beam on the particle. The concept holds potential 
to be used for velocity selection in atomic beams.

This work was supported by the Unity through Knowledge Fund (UKF Grant No. 5/13).
We are grateful to A. Eckardt, J. Radi\'{c}, Th. Gasenzer, and A. Vardi 
for critical reading of the manuscript.

%==============================================================================

\end{document}